\documentclass{webofc}

\usepackage[varg]{txfonts}   
\usepackage{hyperref}
\usepackage{url}
\hypersetup{colorlinks=true,citecolor=blue,urlcolor=blue,linkcolor=blue}
\usepackage{orcidlink}
\begin{document}
%
\title{A heterogeneous and vectorized sequence for the HL-LHC full tracking reconstruction of the CMS experiment}

\author{\firstname{Emmanouil} \lastname{Vourliotis}\inst{1}\orcidlink{0000-0002-2270-0492}\fnsep\thanks{\email{emmanouil.vourliotis@cern.ch}} on behalf of the CMS Collaboration\footnote{Copyright 2026 CERN for the benefit of the CMS Collaboration. Reproduction of this article or parts of it is allowed as specified in the CC-BY-4.0 license.}
}

\institute{University of California San Diego}

\abstract{
This contribution presents the new baseline strategy for the Phase-2 tracking of the CMS experiment for online event reconstruction, and for the main iteration of offline tracking.
This tracking sequence takes advantage of the combination of cutting-edge tracking algorithms that are either optimized for parallel execution on GPUs (Patatrack and LST), or vectorized for efficient CPU performance (mkFit).
Such a combined approach offers an effective solution to deal with the unprecedented computational challenges caused by the large number of simultaneous collisions per bunch crossing expected at the High-Luminosity Large Hadron Collider (HL-LHC).
The proposed combination not only reduces the computational resource requirements but also enhances the physics reach by incorporating displaced tracking and increasingly leveraging machine learning techniques.
}
\maketitle
\section{Introduction}
The High-Luminosity Large Hadron Collider (HL-LHC) promises to deliver an unprecedented amount of high-energy collision data.
This is expected to expand the physics reach of the CMS experiment, enabling precision measurements of the Higgs boson and searches for physics beyond the Standard Model.
However, this scientific potential comes at a significant computing cost.
The steep increase in instantaneous luminosity is the result of the increased number of simultaneous proton-proton collisions (pileup, PU) per bunch crossing, which significantly complicates the event reconstruction. 

A typical HL-LHC event recorded by CMS can contain approximately 200 PU interactions, which would correspond to $\mathcal{O}(10^4)$ charged particle tracks, meant to be reconstructed from $\mathcal{O}(10^5)$ detector hits.
Because the combinatorial complexity of connecting detector hits into tracks scales superlinearly with the hit density, charged particle tracking is historically one of the most computationally expensive steps in event reconstruction.
Without significant improvements, online algorithms operating within the strict computing and timing budgets of the software trigger (High-Level Trigger, HLT) would fail to keep pace with the massive data-taking rates. 

To prevent tracking from becoming a bottleneck, a complete paradigm shift in the software architecture is required.
The CMS Collaboration has been developing a computationally-efficient, highly-parallelized solution for the Phase-2 reconstruction, designed specifically to exploit modern computing techniques, such as CPU vectorization, and modern hardware, such as GPUs.

\section{The CMS Phase-2 tracker upgrade}
\begin{figure}[htbp]
\centering
\includegraphics[width=0.8\textwidth]{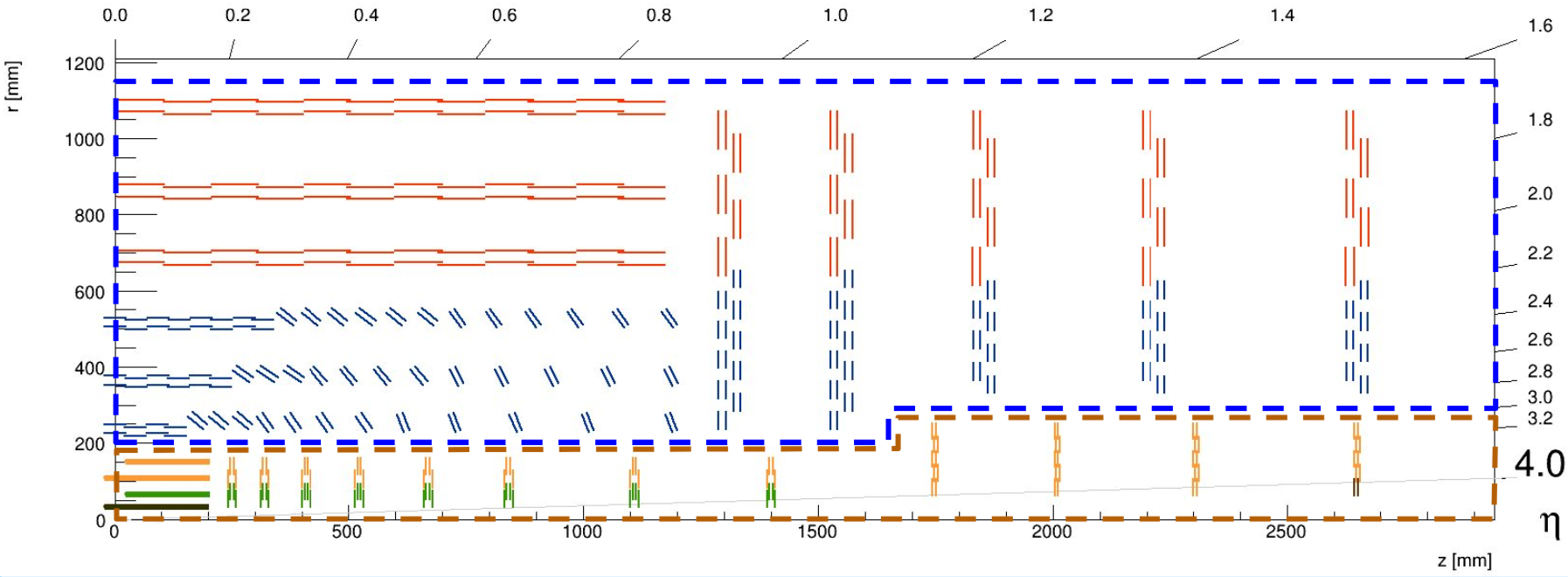}
\caption{Sketch of the Phase-2 CMS tracker with the IT highlighted by an orange frame and the OT highlighted by a blue frame.}
\label{fig:tracker}
\end{figure}

To cope with the increased radiation levels and hit densities of the HL-LHC, the CMS detector is undergoing a major overhaul, known as the ``Phase-2 Upgrade''~\cite{HLTTDR}.
The most critical component for charged particle reconstruction is the new silicon tracker~\cite{TKTDR}, which features finer sensor granularity, expanded geometric coverage up to a pseudorapidity of $|\eta| = 4.0$, and significantly reduced material budget.
The upgraded tracker, schematically shown in Figure~\ref{fig:tracker}, is structurally and functionally divided into two main components: the inner tracker (IT) and the outer tracker (OT).

The IT is composed of densely segmented silicon pixel sensors designed to survive extreme radiation doses.
The IT will feature smaller pixel sizes compared to the current detector, allowing for excellent primary and secondary vertex resolution.
This high granularity is vital for identifying tracks originating from heavy-flavor hadron decays (b tagging) and separating tracks originating from different PU vertices.

The OT introduces a novel concept: $p_\text{T}$-modules. These modules consist of two closely spaced silicon layers (either one macro-pixel and one strip, or two strip sensors).
The front-end electronics of these modules correlate hits in the two layers in real time, forming composite hits, called ``mini-doublets'' (MDs) in this context (also known as ``stubs'').
The MDs can inherently filter out tracks based on their transverse momentum ($p_\text{T}$), significantly reducing the data volume.
While primarily designed to enable tracking at the hardware trigger (Level-1 trigger), the MDs and the OT module geometry provide exciting reconstruction opportunities at the HLT and offline reconstruction, including highly-efficient displaced tracking for long-lived particles.

\section{New algorithms for the CMS Phase-2 tracking: physics and computing performance}
Traditionally, track reconstruction in CMS is performed in distinct sequential steps:

\begin{enumerate}
    \item Track seeding: hits in the tracker are organized in small sets, called ``track seeds'', which serve as a starting point for the reconstruction of full tracks.
    \item Track building: track seeds are extended by adding more hits on the same layers and/or on extra layers, creating ``track candidates''.
    \item Track fitting: track candidates are fitted and the corresponding track parameters are extracted, producing the final tracks.
    Optionally, final tracks undergo a selection, based on quality criteria.
\end{enumerate}

In the following, the physics performance of different tracking sequences is compared.
For this purpose, a reconstructed track is considered matched to a simulated track if more than 75\% of its hits are associated with this simulated track, otherwise the reconstructed track is marked as a fake track.
The tracking efficiency is defined as the fraction of simulated tracks matched to at least one reconstructed track, and the tracking fake rate as the fraction of reconstructed tracks that are fake.
The tracking resolution is defined as the width of the Gaussian curve fitting the distribution of the parameter response.
Only simulated tracks originating from the signal (hard scattering) vertex are used for the efficiency computation, while all simulated tracks are used for the fake rate and resolution computations.

The ``legacy'' Phase-2 configuration executes the track reconstruction steps in multiple iterative passes, with the legacy tracking algorithms, as described in Reference~\cite{TRK}, largely relying on an implementation of a Combinatorial Kalman Filter (CKF) algorithm~\cite{CKF}.
In this legacy configuration, track seeding is tuned to maximize tracking efficiency, which leads to a seeding fake rate as high as 80\%, creating a large timing cost for downstream track building algorithms.
In the HL-LHC environment, this approach consumes approximately 50\% of the entire HLT computing budget.

\begin{figure}[htbp]
\centering
\includegraphics[width=0.7\textwidth]{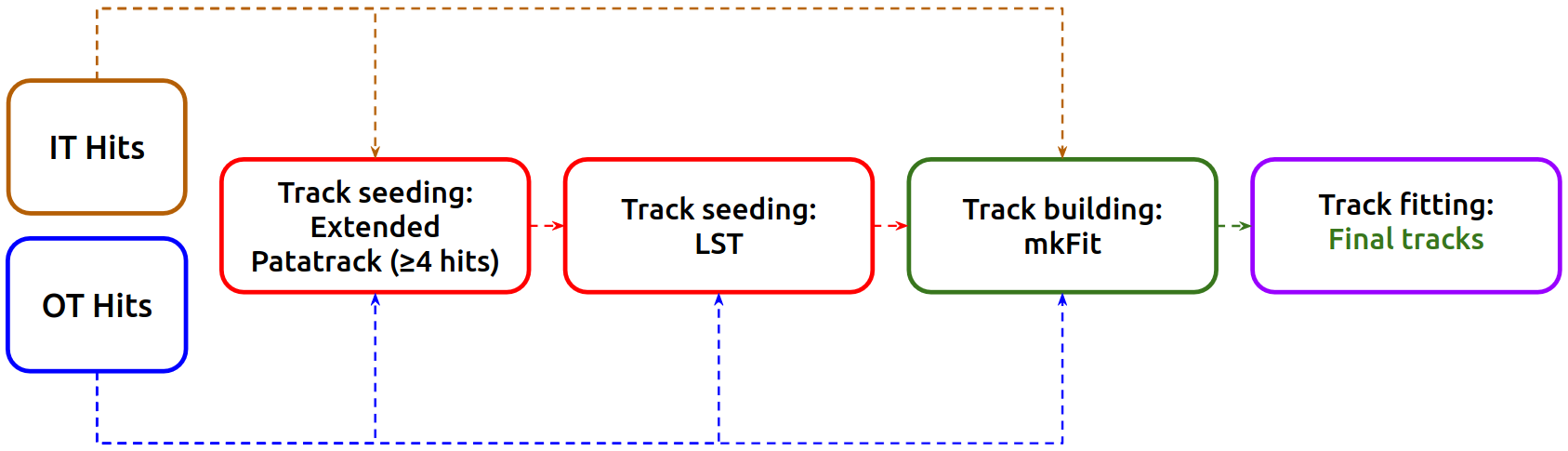}
\caption{A diagram of the new tracking configuration proposed for the CMS Phase-2 reconstruction at HLT.
The arrows indicate the flow of inputs or products from one step to the next.}
\label{fig:TRKSeq}
\end{figure}

To address this issue, the CMS Phase-2 HLT tracking sequence has been fundamentally restructured into a single, highly-optimized iteration~\cite{ThisDPNote}.
This new pipeline utilizes three distinct algorithms designed explicitly for modern, massively-parallel computational architectures:
Extended Patatrack~\cite{Patatrack,ExtPatatrack} and Line Segment Tracking (LST)~\cite{LST,LSTmkFit} for track seeding, and mkFit~\cite{mkFit,LSTmkFit} for track building.
A schematic representation of the new tracking sequence is shown in Figure~\ref{fig:TRKSeq}.

\subsection{Extended Patatrack}
The original Patatrack algorithm was successfully deployed in CMS during the Run 3 of the LHC to perform pixel-only tracking on GPUs.
For Phase-2, this has been upgraded to ``Extended Patatrack'', utilizing hits not only from the IT but also from the pixel-based sensors of the first three barrel layers of the OT.
It is a heterogeneous, hardware-agnostic algorithm written using the Alpaka portability framework~\cite{Alpaka1,Alpaka2,Alpaka3}, allowing the same source code to compile for CPUs, NVIDIA GPUs, and AMD GPUs.

By requiring a minimum of four hits to form a track seed and utilizing a highly-parallelized Cellular Automaton algorithm to connect these hits, Extended Patatrack achieves high purity.
Thus, using the Extended Patatrack algorithm results in a track seeding fake rate less than 10\% (Figure~\ref{fig:seeds}, left), reducing the combinatorial complexity for downstream algorithms.

\subsection{Line Segment Tracking (LST)}
To complement the Extended Patatrack algorithm and recover efficiency for particles that do not leave sufficient hits in the inner layers, such as the decay products of long-lived particles, the LST algorithm is introduced as a track seeding algorithm. 
The LST algorithm is also heterogeneous and hardware-agnostic, implemented in the Alpaka portability framework, but it focuses on the OT.

\begin{figure}[htbp]
\centering
\includegraphics[width=0.40\textwidth]{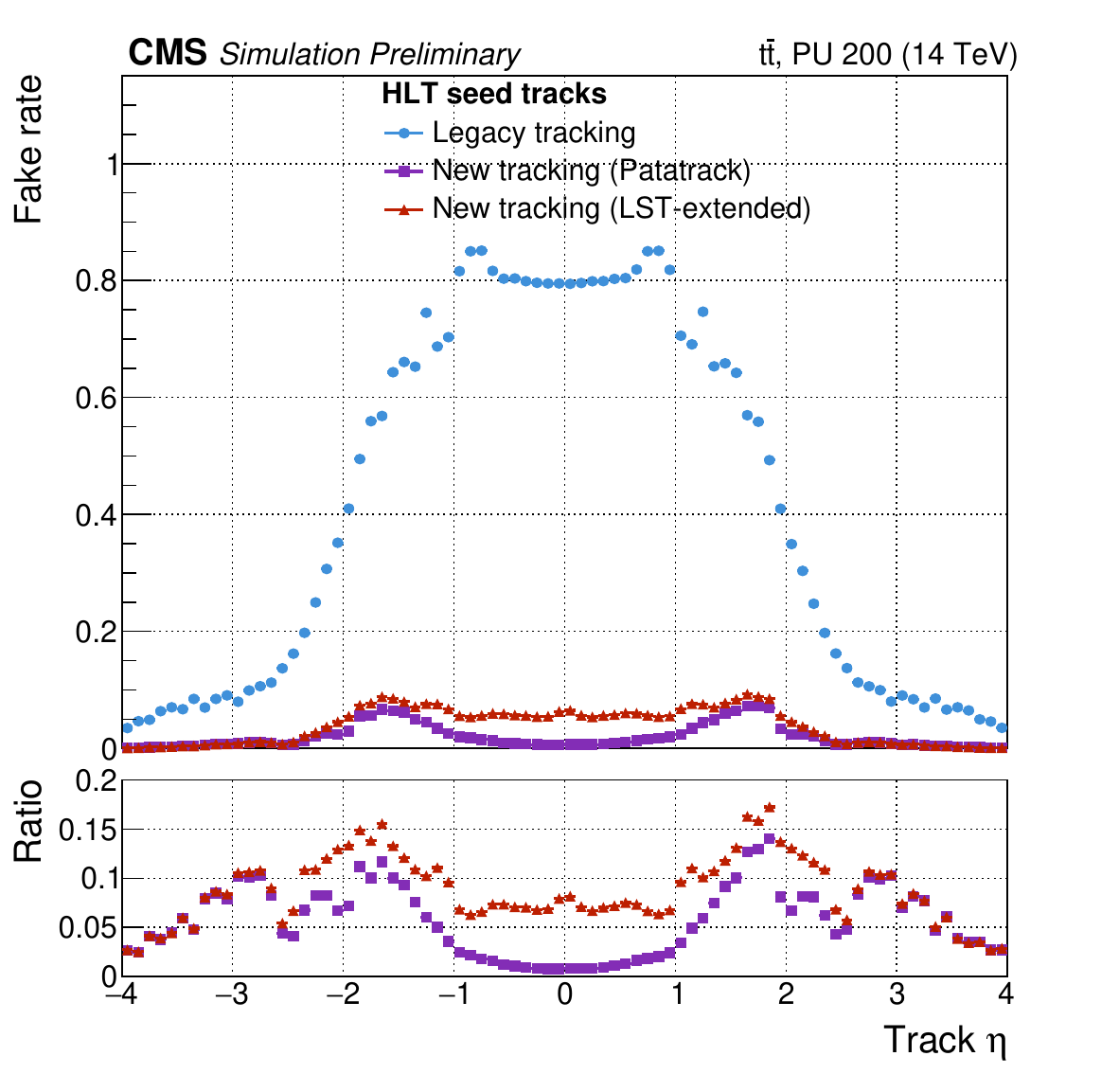}
\includegraphics[width=0.40\textwidth]{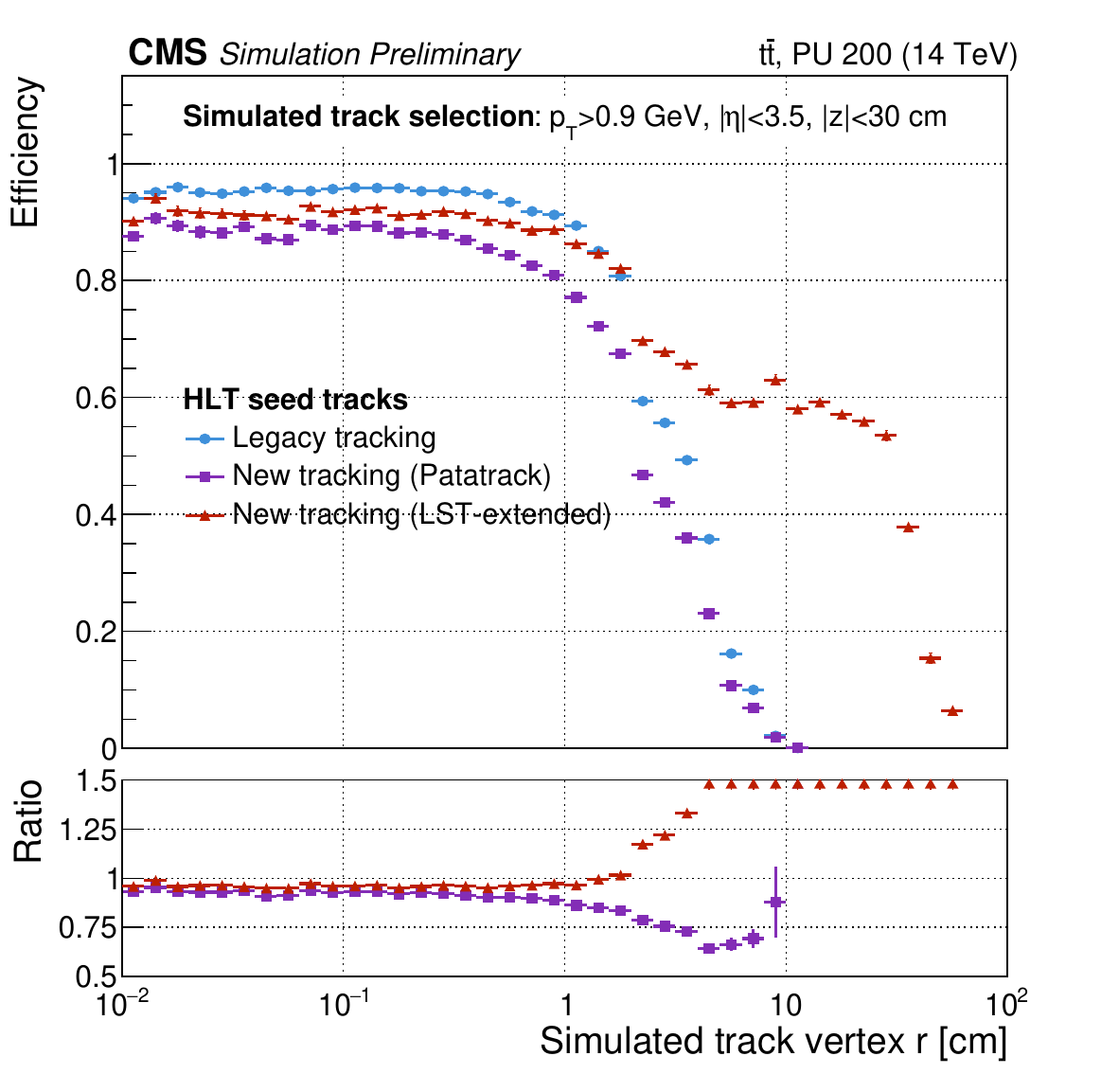}
\caption{Fake rate as a function of the track pseudorapidity $\eta$ (left) and efficiency as a function of the simulated track vertex transverse distance r (right) for seed tracks.
The seed tracks from the legacy track seeding algorithms (blue) are compared with the seed tracks from the Extended Patatrack algorithm (violet) and the LST algorithm (red)~\cite{ThisDPNote}.}
\label{fig:seeds}
\end{figure}

The LST algorithm leverages the unique geometry of the OT $p_\text{T}$-modules to progressively build tracks in parallel.
The algorithm operates hierarchically, building long tracks from shorter ones, applying geometric compatibility criteria and machine learning techniques to filter the combinatorial background.
First, the MDs are created in all OT modules simultaneously.
MDs in adjacent layers are then linked to create ``line segments'' (LSs) that span two layers.
Once created, the LSs are combined to form candidates that cover three (T3s), four (T4s), or five (T5s) layers of the OT.
Finally, the track seeds from the Extended Patatrack algorithm are linked to compatible OT T3 and T5 candidates, creating track seeds that span the full tracker (pT3s and pT5s).
A combination of duplicate-cleaned pT5, pT3, T5, T4 and unused Extended Patatrack track seeds is propagated in this priority order downstream for track building.
In this way, LST enhances the overall efficiency for track seeding on top of Extended Patatrack at a small fake rate cost (Figure~\ref{fig:seeds}).
Crucially, LST adds significant displaced tracking acceptance, being sensitive to transverse displacements up to 60~cm from the beamline.

\subsection{mkFit}
The mkFit algorithm is a modernized, highly parallelized and vectorized implementation of the Kalman filter logic.
It achieves vectorization by utilizing Matriplex, a custom matrix library optimized for Single Instruction, Multiple Data operations relevant for tracking~\cite{mkFit}.
It parallelizes track reconstruction at multiple levels, across different events, detector regions, and groups of track seeds, and manages to minimize memory usage and access latency by utilizing a simplified description of the detector geometry.
After its successful deployment in CMS for track building during the Run 3 of the LHC, it has been adapted to run efficiently for the CMS Phase-2 tracker and HL-LHC conditions, performing track building on top of the inputs from Extended Patatrack and LST algorithms.

\begin{figure}[htbp]
\centering
\includegraphics[width=0.40\textwidth]{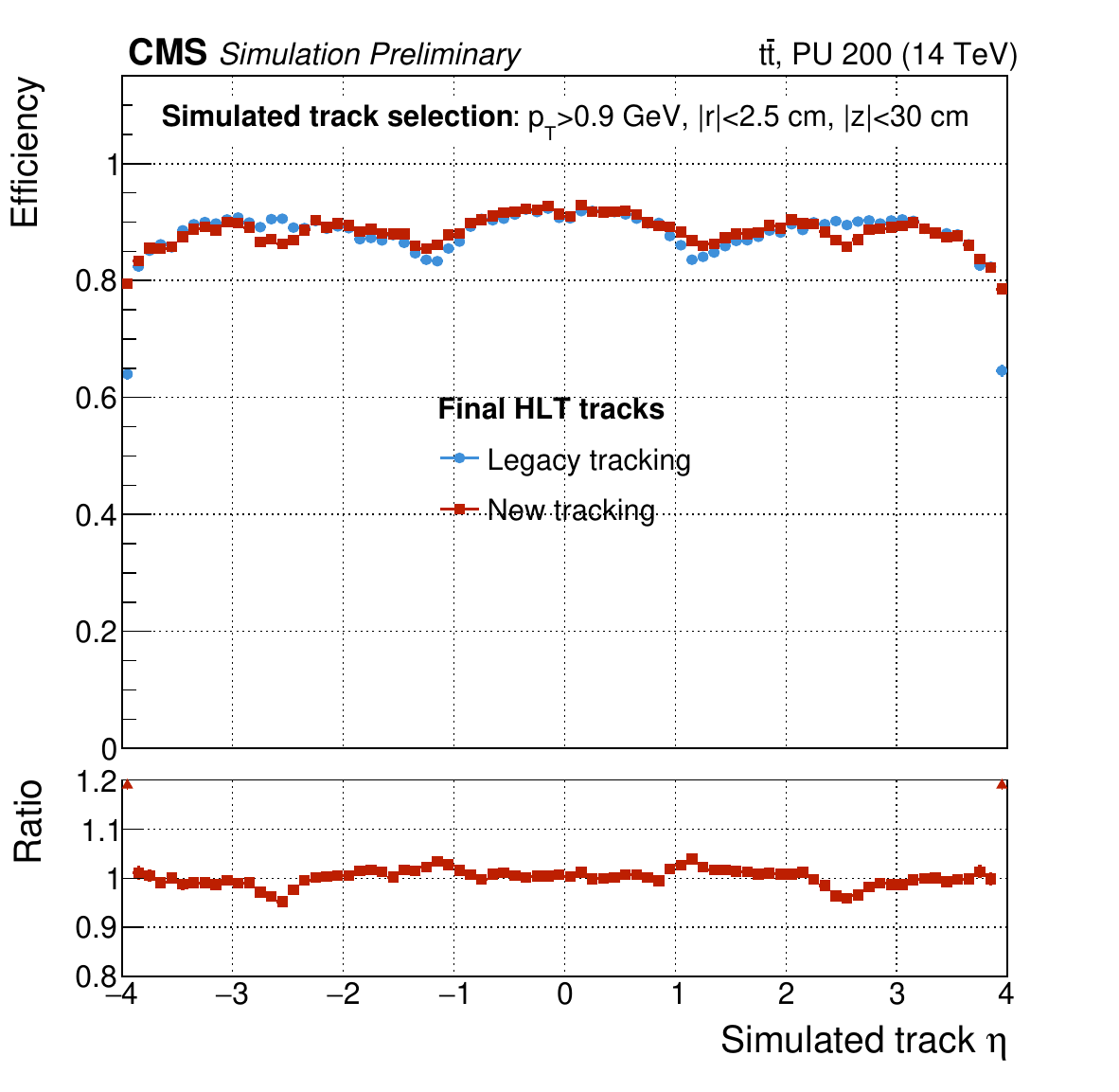}
\includegraphics[width=0.40\textwidth]{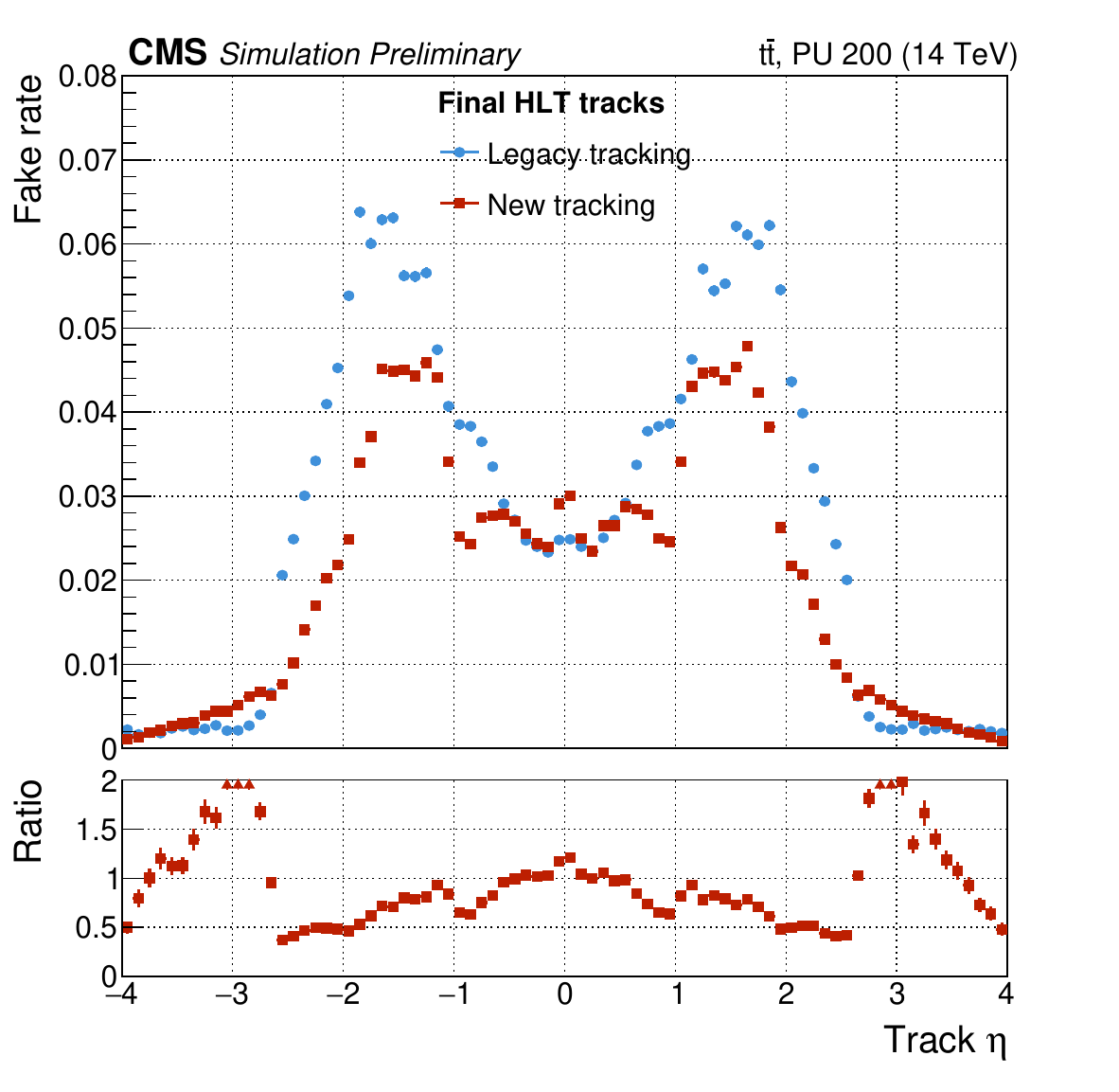}
\caption{Efficiency (left) and fake rate (right) for final tracks as a function of the simulated and reconstructed track pseudorapidity $\eta$, respectively.
The final tracks from the legacy tracking sequence (blue) are compared with the final tracks from the new tracking sequence (red)~\cite{ThisDPNote}.}
\label{fig:tracks_eff_FR}
\end{figure}

\begin{figure}[htbp]
\centering
\includegraphics[width=0.40\textwidth]{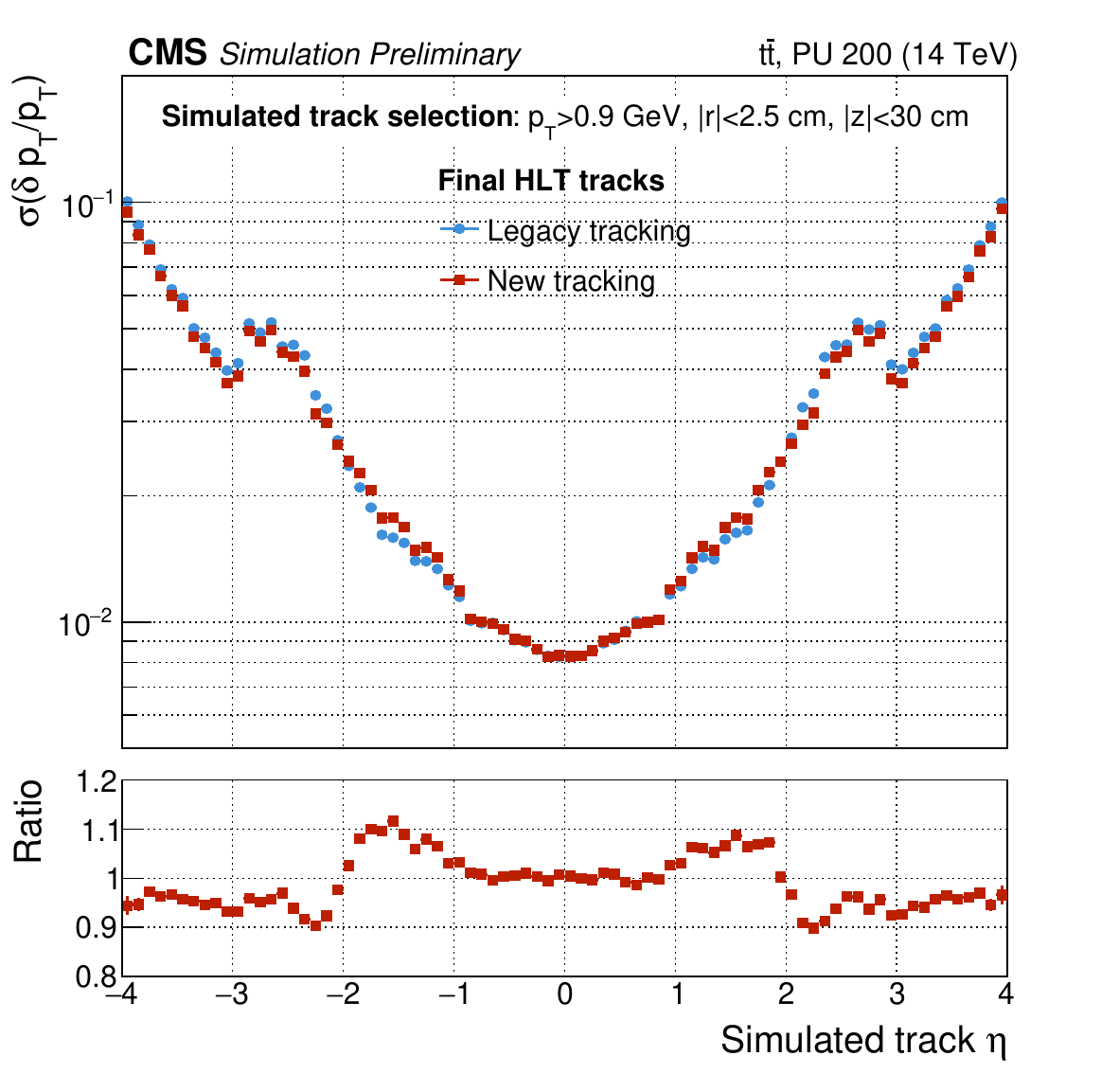}
\includegraphics[width=0.40\textwidth]{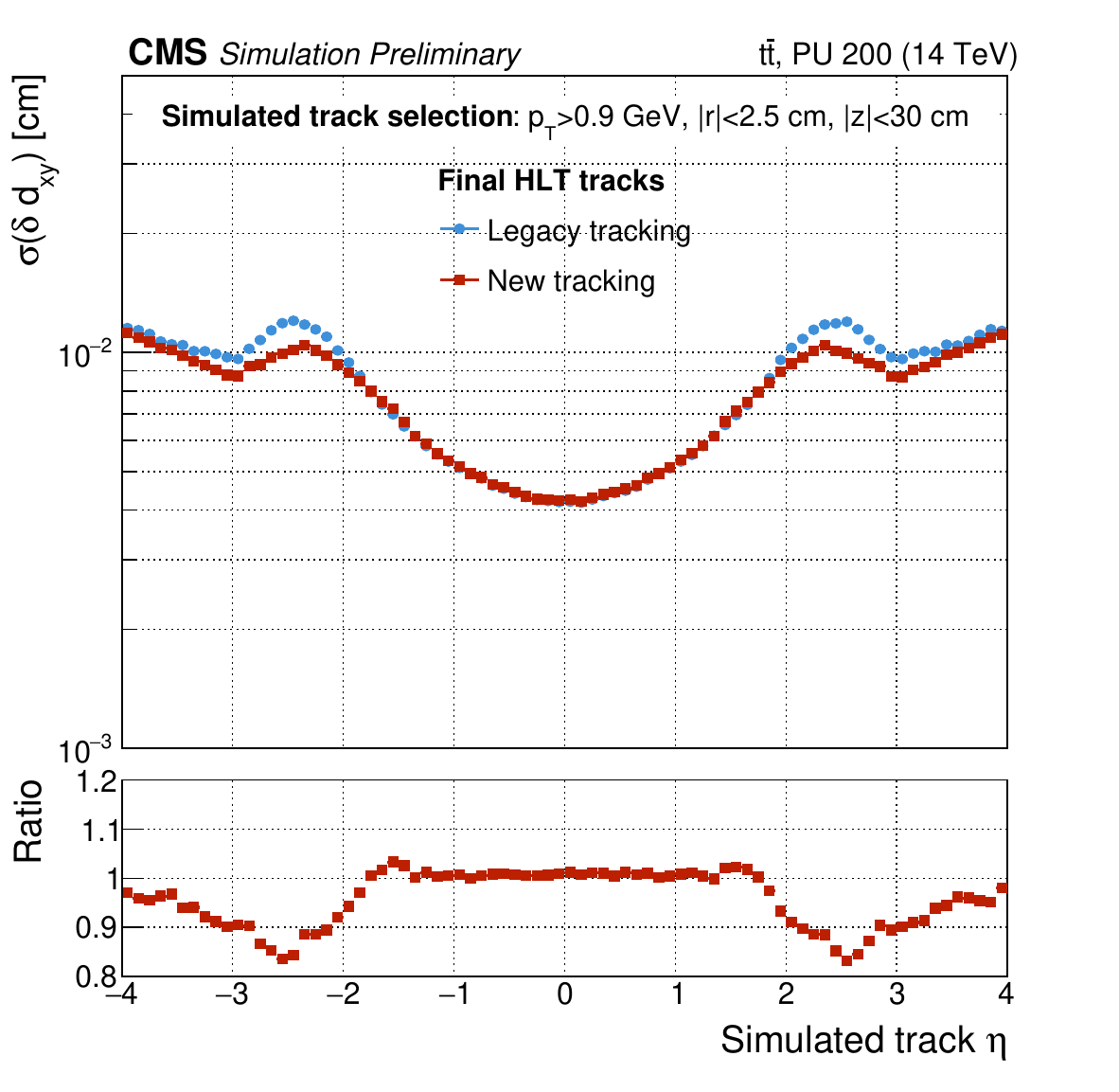}
\caption{Tracking $p_\text{T}$ (left) and $d_\text{xy}$ (right) resolutions for final tracks as a function of the simulated track pseudorapidity $\eta$.
The final tracks from the legacy tracking sequence (blue) are compared with the final tracks from the new tracking sequence (red)~\cite{ThisDPNote}.}
\label{fig:tracks_res}
\end{figure}

The mkFit track building step retains a tracking efficiency comparable to the legacy configuration, while exhibiting a significantly reduced fake rate across the majority of the kinematic phase space (Figure~\ref{fig:tracks_eff_FR}).
In terms of track parameter resolution, mkFit produces tracks with equivalent $p_\text{T}$ and transverse impact parameter ($d_\text{xy}$) resolutions for most of the phase space, with visible improvements in the forward endcap regions of the detector (Figure~\ref{fig:tracks_res}).

\subsection{improvements in computing performance}
The CMS computing target for Run 4 of the Phase-2 HLT is a 50-50 split of processing power between CPUs and GPUs.
Such a setup can be achieved by running the HLT reconstruction with 8 parallel jobs, each with 16 threads and 16 streams, on one AMD EPYC “Milan” 7763 CPU with 128 cores (64 physical cores × 2 logical cores) and 2 NVIDIA T4 GPUs.
Measurements of the HLT reconstruction time with this computing configuration are performed to highlight the improvements accomplished with the new tracking sequence compared to the legacy one.

\begin{figure}[htbp]
\centering
\sidecaption
\includegraphics[width=0.55\textwidth]{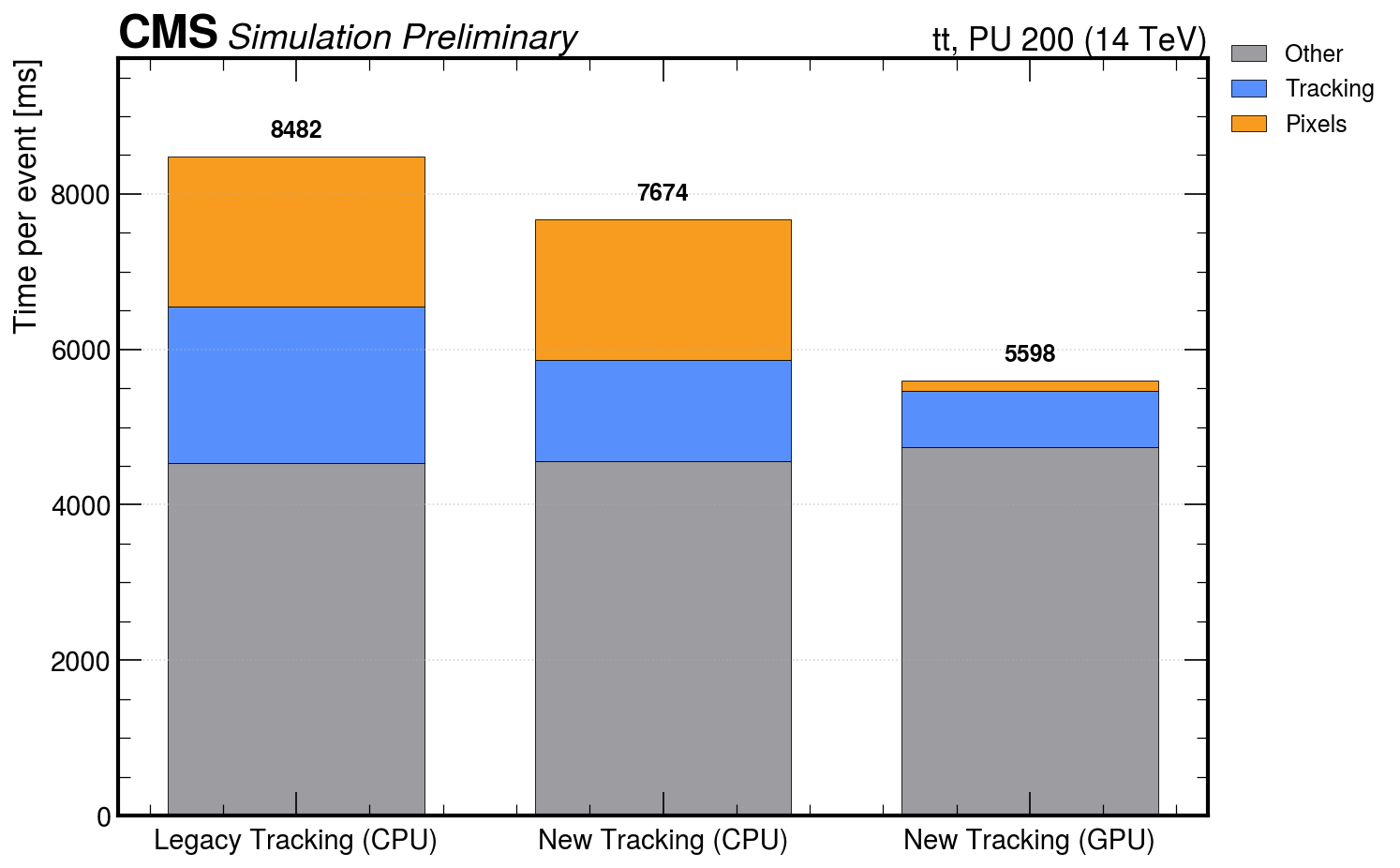}
\caption{Time needed by a logical core to perform the HLT reconstruction on a $t\bar{t}$ event at PU 200. The left column executes the legacy, CPU-only tracking sequence, while the middle column and right columns run the new tracking sequence without (CPU-only) and with offloading to GPU, respectively. The pixel seeding (orange) and the rest of tracking (blue) reconstruction are shown as separate contributions, while the rest of the reconstruction (gray) does not change across columns~\cite{ThisDPNote}.}
\label{fig:timing}
\end{figure}

As shown in Figure~\ref{fig:timing}, running the new tracking configuration on CPU results in a 9\% reduction of the total HLT processing time compared to the CPU-only legacy configuration.
This is achieved thanks to the optimizations implemented in the new algorithms and the removal of the multi-iteration tracking approach.
When the new tracking sequence is offloaded to GPUs, the overall HLT reconstruction timing is reduced further, by 33\% with respect to the CPU-only legacy sequence.

Looking closer at individual algorithm performances, replacing the legacy pixel seeding algorithm with the Extended Patatrack algorithm reduces the ``Pixels'' timing by 6\% on CPU-only and by 93\% with GPU offloading.
Replacing the legacy CKF building algorithm with the combined LST and mkFit algorithms yields a 35\% and 64\% reduction in the track building timing on CPU-only and with GPU offloading, respectively.
The different timing reductions are summarized in Table~\ref{tab:timing}.

\begin{table}[htbp]
    \centering
    \caption{Fractional decrease in reconstruction time compared to legacy tracking sequence~\cite{ThisDPNote}.}
    \label{tab:timing}
    \begin{tabular}{l|cc}
        & New tracking (CPU-only) & New tracking (GPU offloading) \\
        \hline
        Pixels   & 6\%  & 93\% \\
        Tracking & 35\% & 64\% \\
        Overall  & 9\%  & 33\% \\
    \end{tabular}
\end{table}

\section{Conclusion and outlook}
The computationally extreme conditions of the HL-LHC require highly innovative software solutions.
Traditional, sequential tracking algorithms are no longer viable, especially under the strict computing constraints of the Phase-2 HLT.
The newly developed tracking sequence, comprising the Extended Patatrack, the LST, and the mkFit algorithms, represents a major advancement in the CMS event reconstruction performance.

Together, these algorithms deliver a speed-up of up to 33\% for the full HLT event reconstruction, while providing equivalent tracking efficiency, significantly lower fake rates, and substantially extended acceptance for displaced tracks.
Furthermore, the innovations and optimizations originally developed to satisfy HLT latency requirements are actively being adapted for offline track reconstruction.
This unified approach will ensure an efficient and consistent track reconstruction for the operation of CMS at the HL-LHC.\\

\begin{acknowledgement}
This work was supported by the National Science Foundation under Cooperative Agreements OAC-1836650 and PHY-2323298.
\end{acknowledgement}

\end{document}